\documentclass[12pt]{article}
\usepackage{graphicx}
\newfont{\sfb}{cmssbx10 scaled 1400}
\newfont{\bigsf}{cmssbx10 scaled 1600}
\renewcommand{\thefootnote}{\fnsymbol{footnote}}

\begin{document}
\begin{flushright}
RCNP--Th02021\\
FUT--02--01\\
December~~~2002
\end{flushright}
\begin{center}
{\baselineskip 22pt
{\LARGE\bf Polarized light-flavor antiquarks from
Drell-Yan processes of
$h+\vec{N}\to\vec{\ell^{\pm}} + \ell^{\mp} + X$}\\
}
\vspace{2.0em}
{
\renewcommand{\thefootnote}{\fnsymbol{footnote}}
H. Kitagawa\footnote[2]{E-mail: khisashi@post.kek.jp}\\
\vspace{0.8em}
{\it Institute of Particle and Nuclear Studies,
High Energy Accelerator Research Organization (KEK),}\\
{\it Tsukuba, Ibaragi 305--0801, Japan}\\

\vspace{1em}

Y. Sakemi\footnote[9]{E-mail: sakemi@rcnp.osaka-u.ac.jp}\\
\vspace{0.8em}
{\it Research Center for Nuclear Physics,}\\
{\it Osaka University, Ibaraki, Osaka 567--0047, Japan}\\
\vspace{1em}
and\\
\vspace{1em}
T. Yamanishi\footnote[3]{E-mail: yamanisi@ccmails.fukui-ut.ac.jp}\\
\vspace{0.8em}
{\it Department of Management Science,}\\
{\it Fukui University of Technology, Gakuen, Fukui 910--8505, Japan}\\
\vspace{3.5em}
}

{\bf Abstract}
\end{center}

\baselineskip=22pt

We propose a formula to determine the first moment of difference between
 the polarized $\bar u$-
 and $\bar d$-quarks in the nucleon, {\it i.e.} $\Delta\bar u-\Delta \bar d$
 from the Drell-Yan processes in collisions of unpolarized
 hadrons with longitudinally polarized nucleons by measuring
 outgoing lepton helicities.
As coefficients in the differential cross
 section depend on the
 $u$- and $d$-quark numbers in the unpolarized hadron beam,
 the difference $\Delta\bar u-\Delta\bar d$ can be independently tested by
 changing the hadron beam.
Moreover, a formula for estimating the $K$-factor in Drell-Yan
 processes is also suggested.

\vspace{1.0em}
\noindent
PACS number(s): 13.88.+e, 13.85.Ni, 14.65.Bt

\vfill\eject

\baselineskip 22pt
\noindent

Several experiments have been devoted to investigate the nucleon
 structures.
In the large momentum transfer region,
 the nucleon (generally hadron) can be regarded as composed of almost free
 point-like constituents with spin $1/2$, $i.e.$ partons.
It seems quite natural to identify partons as
(almost) massless quarks.
Then, how do quarks construct the nucleon?
Inclusive deep inelastic scatterings (inclusive DIS) off lepton beams on
 nucleon targets have revealed some combinations
 of valence and sea quark distributions in the nucleon.
Traditionally, the light-antiquark distributions, $\bar u(x)$
 and $\bar d(x)$, have been taken to be
 flavor symmetric in phenomenological analyses for structure functions
 of nucleons from a point of view, which the strong interaction
 is independent of the quark flavor for light quark-pair creations
 from gluons.
However, the asymmetry of $\bar u(x)$ and $\bar d(x)$ has been predicted
 \cite{Field}, and in 1991 the NM Collaboration (NMC) at CERN
 has reported on the violation of the Gottfried sum rule\cite{Gottfried}
 from the unpolarized structure functions of
 the proton and the neutron\cite{NMC91}.
This implied a significant flavor asymmetry for
 the unpolarized light-antiquark distributions,
 $\bar u(x)\neq\bar d(x)$.
Furthermore, this asymmetry was confirmed independently by
 the NA51 at CERN\cite{NA51} and the E866 at FNAL\cite{Hawker,Peng}
 through the Drell-Yan (DY) processes with proton beam and proton and
 deuteron targets
 at rapidity $y=0$ and for a large $x_F$ ($i.e.$ small $x$) region.
At present, from these experimental results
 several approaches such as chiral quark model, Skyrme
 model, Pauli blocking effects, etc., are proposed to understand
 light-flavor antiquark asymmetry\cite{Kumano}.
However, the discussions are still under going.
Since the strong interaction does not depend on the quark flavor for
 perturbative
 QCD, the mechanism for its flavor asymmetry
 may be addressed to nonperturbative QCD effect\cite{Diakonov}.
Therefore, the studies on its mechanism are related to disclose how
 quarks build
 up the hadron, and are a challenging subject in particle and nuclear
 physics.

We also come to mind a question whether the longitudinally polarized
 light-antiquark distributions have this asymmetry or not.
Unfortunately,  we have no idea to
 estimate the first moment of $\Delta\bar u - \Delta\bar d$
 such as the Gottfried sum rule for the unpolarized light-antiquark
 flavor asymmetry.
It makes the analysis on the light-flavor antiquark polarization difficult.
The experimental data for the longitudinally polarized semi--inclusive
 DIS \cite{SMC98}-\cite{Funk} have been reanalyzed recently for
 $\Delta\bar u(x)$ and $\Delta\bar d(x)$ \cite{Morii}.
The result implied the asymmetry between $\Delta\bar u(x)$
 and $\Delta\bar d(x)$, but had ambiguities because of
 insufficient information on the fragmentation functions and of
 statistical error.
At present we cannot derive any conclusion on
 $\Delta\bar u(x)$ and $\Delta\bar d(x)$  from data,
though several models with this asymmetry were
 proposed\cite{Dorokov,Gluck}.
Thus, it is important to find a formula for the first moment
 of $\Delta\bar u - \Delta\bar d$, and to measure it directly without
 ambiguites.

So far, several approches for polarized light-flavor antiquark
 distributions in the DY process\cite{Gluck,DY}
 and the weak boson production\cite{Wp} have been
 proposed, but
 these processes can only probe $\Delta\bar u$ and $\Delta\bar d$ in a
 limited kinematic region.

Here we propose a formula for the first moment
 of $\Delta\bar u - \Delta\bar d$
 from the DY process with longitudinally polarized nucleon targets
 and unpolarized hadron beams by measuring the helicity of one of the
 produced pair lepton.

Let us consider a process of
 $h+\vec{N}\to\vec{\ell^{\pm}} + \ell^{\mp} + X$.
The spin-dependent cross section is 
 defined by\cite{Soffer,Chen}
 \begin{equation}
  \frac{d^3\Delta\sigma}{dx_1 dx_2 d\cos\theta_{\mu}}
   \equiv\frac{d^3\sigma_{++}}{dx_1 dx_2 d\cos\theta_{\mu}}-
   \frac{d^3\sigma_{+-}}{dx_1 dx_2 d\cos\theta_{\mu}}~,
   \label{eqn:E2}
 \end{equation}
 where the subscript ($+-$) 
 means the target nucleon helicity and that of one of the outgoing
 leptons are parallel and antiparallel,
 respectively.
This cross section
 is expressed as
 \begin{eqnarray}
  \frac{d^3\Delta\sigma}{dx_1 dx_2 d\cos\theta_{\mu}}&=&
   K_{pol}~\frac{d\Delta\hat\sigma}{d\cos\theta_{\mu}}\sum_i
   e_i^2\left [ \bar q_i^h(x_1, Q^2) \Delta q_i^N(x_2, Q^2)+
  q_i^h(x_1, Q^2)\Delta\bar q_i^N(x_2, Q^2)
  ~\right ] \nonumber\\
  &\equiv&K_{pol}~\frac{d\Delta\hat\sigma}{d\cos\theta_{\mu}}
   \Delta P^{hN}(x_1, x_2, Q^2)~,
   \label{eqn:E1}
 \end{eqnarray}
 with $x_1\equiv x_{beam}$ and $x_2\equiv x_{target}$
 in the leading order (LO) of QCD.
$K_{pol}$, $\Delta\stackrel{\scriptscriptstyle(-)}{q_i^N}$ and
 $\stackrel{\scriptscriptstyle(-)}{q_i^h}$ in
 eq.(\ref{eqn:E1}) are
 the $K$-factor of this DY process,
 the polarized quark (antiquark) and the unpolarized quark (antiquark)
 distributions with flavor $i$, respectively.
Also $\theta_{\mu}$ is the lepton production angle in the center-of-mass
 frame of $hN$ collisions.
In the LO QCD, the differential cross sections of the subprocess
 in eq.(\ref{eqn:E1}) is given by\cite{Soffer}
 \begin{eqnarray}
  & &\frac{d\Delta\hat\sigma}{d\cos\theta_{\mu}}=\mp~\Delta f(x_1, x_2, Q^2,
  \cos\theta_{\mu})~,
  \label{eqn:E3}\\
  & &\Delta f(x_1, x_2, Q^2, \cos\theta_{\mu})=\frac{\pi\alpha^2}{Q^4}
   \frac{8 x_1^2 x_2^2 p_N^2}{\left\{(x_1+x_2)-(x_1-x_2)\cos\theta_{\mu}\right\}^3}
   \left\{x_1^2(1-\cos\theta_{\mu})^2-x_2^2(1+\cos\theta_{\mu})^2\right\}~,
 \nonumber
 \end{eqnarray}
 with $Q^2=x_1x_2s$ and $p_N$ being the nucleon momentum.
Here the initial $-$ and $+$ signs refer to
 $\Delta q~\bar q\to\vec\ell^+\ell^-$ and
 $\Delta\bar q~q\to\vec\ell^+\ell^-$
 for measuring the positive lepton helicity, respectively.
For negative lepton, $-$ and $+$ refer to
 $\Delta\bar q~q\to\ell^+\vec\ell^-$ and
 $\Delta q~\bar q\to\ell^+\vec\ell^-$, respectively.

When the target is the longitudinally polarized proton target,
 $\Delta P^{hN}(x_1, x_2, Q^2)$ in eq.(\ref{eqn:E1}) can be written by
 \begin{eqnarray}
  \Delta P^{hp}(x_1, x_2, Q^2)&=&~~
   \frac{4}{9}\left [\bar u^h(x_1, Q^2)~\Delta u^p(x_2, Q^2)
       -u^h(x_1, Q^2)~\Delta\bar u^p(x_2, Q^2)\right ]\nonumber\\
  & &+\frac{1}{9}\left [\bar d^h(x_1, Q^2)~\Delta d^p(x_2, Q^2)
   -d^h(x_1, Q^2)~\Delta\bar d^p(x_2, Q^2)\right ]\nonumber\\
  & &+\frac{1}{9}\left [\bar s^h(x_1, Q^2)~\Delta s^p(x_2, Q^2)
   -s^h(x_1, Q^2)~\Delta\bar s^p(x_1, Q^2)\right ]\nonumber\\
  & & + {\rm ( contributions~~from~~heavy~~quark~~distributions ) }~.
 \label{eqn:E4}
 \end{eqnarray}
Similary, $\Delta P^{hn}(x_1, x_2, Q^2)$ for the neutron target
 can be also obtained.
Assuming the isospin symmetry for the target nucleon, one has an
 interesting equation such as,
 \begin{eqnarray}
  & &\frac{d^3\Delta\sigma^{hp}}{dx_1 dx_2 d\cos\theta_{\mu}}
   -\frac{d^3\Delta\sigma^{hn}}{dx_1 dx_2 d\cos\theta_{\mu}}
   =K_{pol}~\frac{d\Delta\hat\sigma
   } {d\cos\theta_{\mu}}\left\{\Delta P^{hp}(x_1, x_2, Q^2)-
       \Delta P^{hn}(x_1, x_2, Q^2)\right\}\nonumber\\
  & &=\mp~K_{pol}~\Delta f(x_1, x_2, Q^2, \cos\theta_{\mu})\nonumber\\
  & &\times\left[\frac{1}{9}\left [\Delta u_v^p(x_2, Q^2)
     -\Delta d_v^p(x_2,Q^2)
     +2\left\{\Delta\bar u^p(x_2, Q^2)-\Delta\bar d^p(x_2,
Q^2)\right\}\right ]
     \left\{4\bar u^h(x_1, Q^2)-\bar d^h(x_1, Q^2)\right\}\right
.\nonumber\\
  & &~-\frac{1}{9}\left\{\Delta\bar u^p(x_2, Q^2)-
   \Delta\bar d^p(x_2, Q^2)\right\}
   \left[4u_v^h(x_1, Q^2)-d_v^h(x_1, Q^2)+2
    \left\{4\bar u^h(x_1, Q^2)-\bar d^h(x_1, Q^2)\right\}
    \left .\frac{}{} \right ]\right ]~,\nonumber\\
  & &
  \label{eqn:E5}
 \end{eqnarray}
where the initial $-$ and $+$ signs in the right-hand side of above equation
correspond to the measurement of the positively charged lepton and
negatively one, respectively.
%
Integrating over $x_1$ and $x_2$ on eq.(\ref{eqn:E5}),
 we obtain the following relation
 \begin{eqnarray}
  & &\int^1_0\int^1_0\frac{
     d^3\Delta\sigma^{hp}/dx_1 dx_2 d\cos\theta_{\mu}
    - d^3\Delta\sigma^{hn}/dx_1 dx_2 d\cos\theta_{\mu}}
   {K_{pol}~\Delta f(x_1, x_2, Q^2, \cos\theta_{\mu})}dx_1dx_2\nonumber\\
  & &=\frac{1}{9}\left [\Delta u_v^p(Q^2)-\Delta d_v^p(Q^2)+
    2\left\{\Delta\bar u^p(Q^2)-\Delta\bar d^p(Q^2)\right\}\right ]
    \left\{4~\bar u^h(Q^2)-\bar d^h(Q^2)\right\}\nonumber\\
  & &~-\frac{1}{9}\left\{\Delta\bar u^p(Q^2)-\Delta\bar d^p(Q^2)\right\}
   \left [4~u_v^h-d_v^h+2\left\{ 4~\bar u^h(Q^2)-\bar
d^h(Q^2)\right\}\right ]
   \nonumber\\
  & &=\frac{1}{9}\left |\frac{g_A}{g_V}\right |
   \left\{4~\bar u^h(Q^2)-\bar d^h(Q^2)\right\}-\frac{1}{9}
   \left\{\Delta\bar u^p(Q^2)-\Delta\bar d^p(Q^2)\right\}
   \left [4~u_v^h-d_v^h+2\left\{ 4~\bar u^h(Q^2)-\bar
d^h(Q^2)\right\}\right ]\nonumber\\
  & &
  \label{eqn:E6}
 \end{eqnarray}
 for the measurement of the $\ell^-$ helicity.
In the eq.(\ref{eqn:E6}),
 $g_A$ and $g_V$ are the nucleon axial and vector coupling constants,
 respectively,
 and $u_v^h$ and $d_v^h$ are numbers of the valence
 $u$- and $d$-quarks in the beam particle $h$, respectivley.
Here, we drop the label $Q^2$,
 since the valence quark numbers in the hadron are independent of $Q^2$.
Therefore,
 appraising 
 $4\bar u^h(Q^2)-\bar d^h(Q^2)$ in the unpolarized beam hadron $h$
 and the $K$-factor of this polarized DY process,
 we get information on the first moment of
 $\Delta\bar u^p-\Delta\bar d^p$
 from the cross sections for $h + \{\vec{p} \mbox{\rm\ \ and\ \ }
 \vec{n} \}\to \vec{\ell^{\pm}} + \ell^{\mp} + X$ with wide ranges of
 both $x_1$ and $x_2$.

If the polarized light-flavor antiquark distribution is symmetric,
{\it i.e.} $\Delta\bar u^p-\Delta\bar d^p=0$,
 the right-hand side of eq.(\ref{eqn:E6})
 reduces to $1/9|g_A/g_V|\{4\bar u^h(Q^2)-\bar d^h(Q^2)\}$.
For instance, choosing the proton as the unpolarized hadron beam
  $h$ and taking $x_{1~min}=10^{-5}$ and $Q^2=4$GeV$^2$, it becomes
  $1/9\cdot 1.267\cdot 0.14\times 10^2$$=19.7$ with $0.141\times 10^2$
  or $0.139\times 10^2$ for $4\bar u^p(Q^2)-\bar d^p(Q^2)$ by
  the parametrization of GRV98LO\cite{GRV98} or MRST98LO\cite{MRST98},
  respectively.
Also, the difference between the spin-dependent
 differential cross sections of $pp$ and $pn$ collisions as a function of
 $x_2$ for
 several $x_1$ values is shown in fig.1 by taking $K_{pol}=1.8$.
We use AAC\cite{AAC} with $\Delta\bar u=\Delta\bar d $ and GRV98LO
 parametrizations as polarized and unpolarized distribution functions,
 respectively.
Therefore, we can conclude that the behavior of $\Delta\bar u$ and
 $\Delta\bar d$ in the
 nucleon is asymmetric if we find a discrepancy between the
 measured values and above predicted ones.

For other unpolarized hadron beams, for example charged
 pions, kaons and so on, it is not difficult to estimate
 $4\bar u^h(Q^2)-\bar d^h(Q^2)$ in eq.(\ref{eqn:E6}) from experiments.
The value of $4\bar u^h(Q^2)-\bar d^h(Q^2)$
 in the unpolarized beam hadron $h$ can be obtained
 from data in the same way as above.
With unpolarized nucleon targets,
 the same procedure as eq.(\ref{eqn:E6}) leads to
 \begin{eqnarray}
  & &\int^1_0\int^1_0\frac{d^3\sigma^{hp}/dx_1 dx_2 d\cos\theta_{\mu}
   -d^3\sigma^{hn}/dx_1 dx_2 d\cos\theta_{\mu}}
   {K_{unpol}~d\hat\sigma/d\cos\theta_{\mu}}dx_1dx_2 \nonumber\\
  & &=\frac{1}{9}\left\{\bar u^p(Q^2)-\bar d^p(Q^2)\right\}(4~u_v^h-d_v^h)
   +\frac{1}{9}\left [u_v^p-d_v^p
   +2\left\{\bar u^p(Q^2)-\bar d^p(Q^2)\right\}
   \right ]\left\{4~\bar u^h(Q^2)-\bar d^h(Q^2)\right\}~,\nonumber\\
  & &
  \label{eqn:E6-2}
 \end{eqnarray}
 where $K_{unpol}$ is the $K$-factor of the unpolarized DY process, and
 the differential cross sections of the subprocess is written as
 $$
 \frac{d\hat\sigma}{d\cos\theta_{\mu}}=\frac{\pi\alpha^2}{Q^4}
   \frac{8 x_1^2 x_2^2 p_N^2}{\left\{(x_1+x_2)-(x_1-x_2)\cos\theta_{\mu}\right\}^3}
   \left\{x_1^2(1-\cos\theta_{\mu})^2+x_2^2(1+\cos\theta_{\mu})^2\right\}~.
 $$
Since the value of $\bar u^p(Q^2)-\bar d^p(Q^2)$ in eq.(\ref{eqn:E6-2})
 was obtained by the NMC and other experiments and also has been
 studied intensively,
 one can extract
 $4\bar u^h(Q^2)-\bar d^h(Q^2)$ from the combination of the
 differential cross sections, $d\sigma^{hp}$ and $d\sigma^{hn}$,
 for the unpolarized DY processes.
Using the NMC result, namely
 $\bar u^p(Q^2)-\bar d^p(Q^2)=-0.147\pm0.039$ at
 $Q^2=4$GeV$^2$\cite{NMC91}, and choosing
 the proton as the hadron $h$, we obtain
 \begin{eqnarray}
  & &4~\bar u^h(Q^2)-\bar d^h(Q^2)\nonumber\\
  & &=\left[\int^1_0\int^1_0\frac{
       d^3\sigma^{hp}/dx_1 dx_2 d\cos\theta_{\mu}
       -d^3\sigma^{hn}/dx_1 dx_2 d\cos\theta_{\mu}}
       {K_{unpol}~d\hat\sigma/d\cos\theta_{\mu}}
       dx_1dx_2 + 0.114\right] /~0.0784~.
  \label{eqn:E6-3}
 \end{eqnarray}

For $K_{pol}$ and $K_{unpol}$ appeared in eqs.(\ref{eqn:E6})-(\ref{eqn:E6-3}),
 in general, it is important to find an equation where the $K$-factor
 cancels out
 in order to extract physical quantities from DY processes.
However, we do not have enough information derived from such equation.
In order to get absolute values of the physical quantities
 induced 
 from the DY processes,
 the $K$-factor must be estimated exactly.
Using unpolarized antihadron beams together with
 unpolarized hadron beams,
 the mean value of $K_{pol}$ in terms of $x$ for this polarized
 DY process is given
 as follows
 \begin{eqnarray}
  & &\int^1_0\int^1_0\frac{\left\{\frac{d^3\Delta\sigma^{\bar hp}}
    {dx_1 dx_2 d\cos\theta_{\mu}}
    -\frac{d^3\Delta\sigma^{\bar hn}}{dx_1 dx_2 d\cos\theta_{\mu}}\right\}
    -\left\{\frac{d^3\Delta\sigma^{hp}}{dx_1 dx_2 d\cos\theta_{\mu}}
      -\frac{d^3\Delta\sigma^{hn}}{dx_1 dx_2 d\cos\theta_{\mu}}\right\}}
      {\Delta f(x_1, x_2, Q^2, \cos\theta_{\mu})}dx_1dx_2\nonumber\\
  & &=\mp~K_{pol}~\frac{1}{9}\left |\frac{g_A}{g_V}\right |
   \left (4~u_v^h-d_v^h\right )~,
   \label{eqn:E7-1}
 \end{eqnarray}
 with $\mp$ being similar to eq.(\ref{eqn:E5}), 
 where we assume the reflection symmetry along the V-spin axis,
 the isospin symmetry and the charge conjugation invariance for
 the unpolarized beam hadron with the polarized nucleon target.
For $K_{unpol}$ of the unpolarized DY process, we also have
 \begin{eqnarray}
  & &\int^1_0\int^1_0\frac{\left\{\frac{d^3\sigma^{\bar hp}}
    {dx_1 dx_2 d\cos\theta_{\mu}}
    -\frac{d^3\sigma^{\bar hn}}{dx_1 dx_2 d\cos\theta_{\mu}}\right\}
    -\left\{\frac{d^3\sigma^{hp}}{dx_1 dx_2 d\cos\theta_{\mu}}
      -\frac{d^3\sigma^{hn}}{dx_1 dx_2 d\cos\theta_{\mu}}\right\}}
      {d\hat\sigma/d\cos\theta_{\mu}}dx_1dx_2\nonumber\\
  & &=K_{unpol}~\frac{1}{9}\left (u_v^p-d_v^p\right )
   \left (4~u_v^h-d_v^h\right )~.
   \label{eqn:E7-2}
 \end{eqnarray}
Since each term of the right-hand side in eqs.(\ref{eqn:E7-1}) and
 (\ref{eqn:E7-2}) is constant,
 $K_{pol}$ and $K_{unpol}$ can be evaluated from the relevant
 differential cross sections.

Thus, measuring the $K$-factors by using unpolarized
 hadron and antihadron beams, and evaluating
 $4\bar u^h(Q^2)-\bar d^h(Q^2)$ from the unpolarized DY experiment,
 the polarized DY process for an unpolarized hadron beam and
 a polarized nucleon target allows to provide
 the first moment of $\Delta\bar u^p(Q^2)-\Delta\bar d^p(Q^2)$
 by measuring the helicity of one of the produced pair lepton.
Note that
 it does not require measurements with high precision for
 large $x_1$ and $x_2$ region in order to determine
 the value of $\Delta\bar u^p-\Delta\bar d^p$,
 though differential cross sections for the proton and the neutron
 targets are combined.
Because each differential cross section for large $x_1$ and $x_2$
 is quite small as shown in fig.1,
 a contribution from this region to the integral is small.

%

In summary,
 we have proposed a formula for the difference between the polarized
 light-flavor antiquark density,
 $\Delta\bar u^p-\Delta\bar d^p$, from DY processes.
It is given by a combination of the cross sections
 with an unpolarized hadron beam and a longitudinally polarized proton
 target by measuring one of the produced lepton helicity
 and that with a neutron target.
Then, the formula is described in terms of the neutron $\beta$-decay
 constant and the difference between $\Delta\bar u^p$
 and $\Delta\bar d^p$.
As coefficients of these terms depend on
 $u$- and $d$-quark numbers in the unpolarized beam hadron,
 we can independently get information on the behavior
 of $\Delta\bar u^p$ and $\Delta\bar d^p$
 by changing the beam hadron.

Recently the DY process in the next-to-leading order (NLO) of QCD has been
 discussed and compared with that in the LO,
 though it is the DY process for the longitudinally polarized nucleon and
 the longitudinally polarized nucleon collisions\cite{Gluck}.
The relevant differential cross section has an additional
 term including contributions from gluons.
However, in the difference $d\Delta\sigma^{hp}-d\Delta\sigma^{hn}$
 the contibutions from gluons will be cancel.
Accordingly, eq.(\ref{eqn:E6}) is expected to be also kept in the NLO.

Now experiments to solve the nucleon spin problems start
 at Relativestic Heavy Ion Collider (RHIC) in BNL
 with colliding polarized protons at high energy, $\sqrt s = 200$GeV.
Also, Japan Hadron Facility (JHF) is under construction.
It provides 50GeV high intensity proton and antiproton
 beams, and also can produce high intensity charged pi/K meson beams.
One expects that future experiments for processes proposed here
 will be carried at RHIC and/or JHF with the detector measuring
 the helicity of high energy leptons over wide ranges of $x_1$ and $x_2$. 
It is certainly difficult to measure the high energy lepton helicity.
However, the measurement of the helicity of the $\mu^+$ produced
 in charged current interactions has
 already been carried out at high-energy antineutrino
 experiments at CERN-SPS\cite{Jonker}.
The polarimeter used there composed of the marbles for stopping the
 muon, the scintillators for detecting the positron from muon decay and
 the proportional drift tubes for observing the muon track\cite{Jonker}.
This makes the use of the fact that high energy positron is preferentially
 emitted in the direction of the muon spin because of the V-A intereaction
 in muon decay.
Accordingly, it seems possible to do experiments on the processes
 discussed here.
We wish to get new informations on $\Delta\bar u^p-\Delta\bar d^p$
 at RHIC and/or JHF.

\vspace{0.8em}

One of us (T. Y.) thanks T. Morii for enlightening comments,
and K. Kobayakawa for careful reading of this manuscript.

\newpage

\vspace{-2.0cm}
\begin{figure}[h]
\begin{center}
\includegraphics[width=0.8\linewidth]{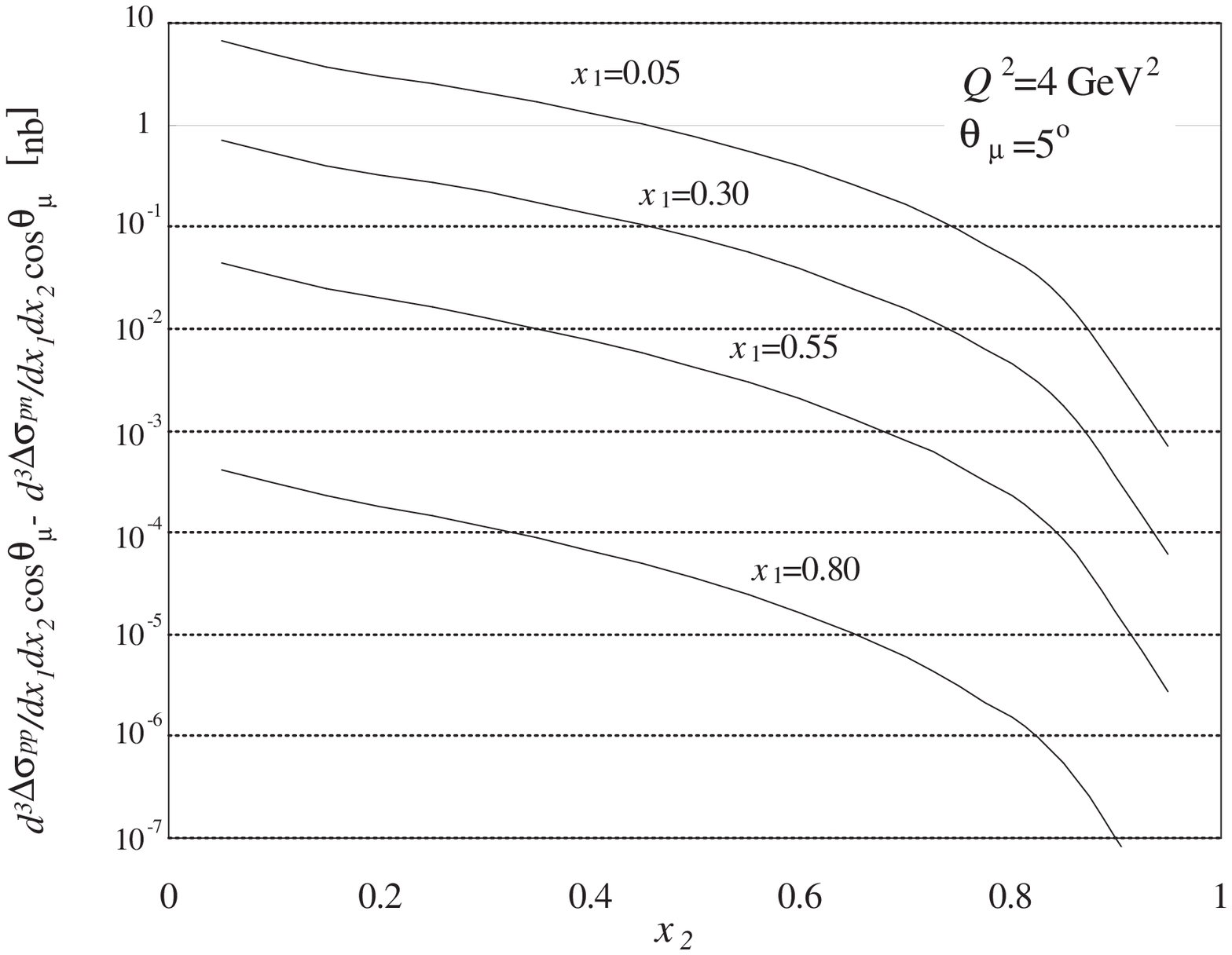}
\caption{The difference between the spin-dependent differential
cross sections of $pp$ and $pn$ collisions for the measurement
of the $\ell^-$ helicity as a function of $x_2$ with
$K_{pol}=1.8$ and $\theta_{\mu}=5^o$.}
\label{fig:fig1}
\end{center}
\end{figure}
\end{document}